\documentclass[10pt]{article}

\usepackage{amsfonts}

\input{tcilatex}

\begin{document}

\begin{center}
\textbf{MODULATIONAL INSTABILITY IN BOSE--EINSTEIN  CONDENSATE IN
OPTICAL SUPERLATTICE}

\bigskip

\vspace{0.5in} Ekaterina A.~Sorokina$^{a}$ \footnote{ electronic
address: sokate@mail.ru} and Andrei I. Maimistov$^{a}$ \footnote{
electronic address: maimistov@pico.miphi.ru}

\vspace{0.2cm} $^{a}$ Department of Solid State Physics, Moscow
Engineering Physics Institute, Kashirskoe sh. 31, Moscow, 115409
Russia
\end{center}

\bigskip

\begin{center}
\textbf{ABSTRACT}
\end{center}

\bigskip

Steady state distribution of the probability amplitudes and the site
population in the one dimensional optical superlattice was found. It
was shown that this solution of the equations which describe the
dynamics of the Bose-Einstein condensate in superlattice is unstable
at the sufficiently high density of the bosons. The expression for
increment of the modulational instability was found on base of the
linear stability analysis. Numerical simulation demonstrates the
evolution of the steady state distributions of bosons into the space
array of the solitary peaks before the chaotic regime generation.


\newpage

\section{Introduction}

Interference of the several plane waves of monochromatic radiation
can form a diffraction pattern in which the electric field strength
periodically varies in space. Resulting periodic system of the
microscopic potentials is designated as optical lattice
\cite{BM,Oplet1,Oplet2}. Bose-Einstein condensates (BECs) trapped in
optical lattice have been studied in sufficient detail
\cite{R1,R2,R3}. The dynamics of atoms in an optical lattice can be
described by two methods. The first method is based on the nonlinear
Schr\"{o}dinger equation with a periodic potential \cite{NLS1,NLS2}.
In papers devoted to investigation of the Bose-Einstein
condensation, this equation is often called the Gross-Pitaevskii
equation \cite{Pita1,Pita2,Leggett}. The second method answering the
tight-binding approximation is based on the Hubbard model, where the
operators of creation and annihilation of fermions are replaced by
the operators satisfying the commutation relations for bosons. The
resulting model is often called the Bose-Hubbard model
\cite{R4,R5,Fialko,R6}.

Recently the optical lattices with two sorts of microscopic
potentials in a unit cell have attracted attention
\cite{R7,R8,R9,R10,R11}. In analogy with theory of solid state
crystals, this periodic system of the microscopic potentials can be
named optical superlattice. Due to the difference of the energy
levels of microscopic potentials quantum tunneling between the
nearest-neighbor sites is absent. Also, at low temperatures
thermoactivated transport of an atom from one site of the
micropotential to the other site is absent. However, the
photo-induced transport of atoms along the optical superlattice is
possible under the condition for the Raman
resonance\cite{R4,R12,R13}. A similar process is known in nonlinear
optics as a coherent population transfer and in solid state physics
of low dimensional systems as coherent transfer of electrons or
excitons in a system of coupled quantum dots.

Under CW electromagnetic radiation the different nonlinear
excitations can propagate in the optical superlattice. It is
important to emphasize that parameters of these excitations can be
controlled by additional radiation, which defines hopping rates
between the adjacent sites.

Frequently the steady state wave motion takes place in various
physical systems. Due to the interplay between dispersive and
nonlinearity effects a week perturbation of the steady state wave
may induce the exponential growth of the perturbation. That
phenomenon is the modulational instability of a steady state wave.
Some times modulational instability results in train of the
soliton-like waves. But it does not always happen.

Modulational instability in BECs in the case of the ordinary optical
lattices was investigated in \cite{R14,R15,R16}. It was shown that
modulational instability is basic mechanism by which solitons are
created in BEC. The kind of solitons (i.e., bright or dark one)
depends on sign of the scattering lengths.

In this paper we consider the dynamics of the site populations for
optical superlattice  with two sorts of microscopic potentials in a
unit cell. The generated Bose-Hubbard model describing this system
in the tight-binding approximation was used in \cite{R12,R13} to
write the system of equations of motion for probability amplitudes
of the site population. These equations are employed as the basis
for present investigation. We found the stationary distribution of
the probability amplitudes. It should be pointed out that this
distribution is not homogeneous one, as opposed to the case of
ordinary optical lattices. The main result is an analytic expression
for the modulational instability increment, which depends on the
site population and the wave number of harmonic weak perturbations.
This expression was obtained on base of the linear stability
analysis. The numerical simulation shows that instability leads to
strongly non-regular pattern.

\section{Model and basic equations}

Let us consider the one-dimensional optical superlattice with two
kinds of sites in the tight-binding approximation ~\cite{R4,R5}. The
microscopic potentials of one depth correspond to sites with an even
number, and microscopic potentials of other depth correspond to
sites with an odd number. Sites labeled by even numbers contain
bosons in the ground state $|g_{a}>$  with the energy
$\varepsilon_{a}$. Sites labeled by odd numbers contain bosons in
the ground state $|g_{b}>$ with the energy $\varepsilon _{b}>
\varepsilon _{a}$. Let the temperature of the system be such that
the higher levels of the microscopic potentials are not populated.
Since the energies of the ground states of neighboring sites differ,
the process of direct tunneling of an atom from one site to another
one can be excluded from the consideration. Let us assume that
biharmonic radiation ($\omega _{1}$ and $\omega _{2}$  are the
frequencies of the carrier waves) acts on the atoms and the
condition for the Raman resonance $ (\varepsilon
_{b}-\varepsilon_{a})\approx \hbar (\omega _{1}-\omega_{2})$ is
fulfilled. In this case, after absorption of the first (second)
photon, the atom will go from the deep (shallow) microscopic
potential to the state of the continuous spectrum, and, after
emission of the second (first) photon, it will return to the state
having the energy $\varepsilon_{b}$ ($\varepsilon_{a}$) and thus
will be brought into a shallow (deep) microscopic potential. Thus,
although tunneling or thermoactivated transport of atoms along the
optical superlattice is absent, their photo-induced transport is
possible.

In classical limit the system of equations describing the
probability amplitudes of the populations for sites $a_{2j}=\langle
\hat{a}_{2j}\rangle$ and $b_{2j+1}= \langle \hat{b}_{2j+1}\rangle$
takes the following form \cite{R12,R13}:
\begin{eqnarray}
i\hbar \frac{\partial }{\partial t}a_{2j} &=& -J_{0}^{\ast
}e^{i\Delta \omega t}+\left( b_{2j-1}+b_{2j+1}\right) +\varepsilon
_{a,2j}a_{2j}+U_{aa}|a_{2j}|^{2}a_{2j}+ \nonumber \\
&& + U_{ab}\left( |b_{2j-1}|^{2}+|b_{2j+1}|^{2}\right) a_{2j},\\
\label{eq1a} i\hbar \frac{\partial }{\partial t}b_{2j+1} &=&
-J_{0}e^{-i\Delta \omega t}+\left( a_{2j}+a_{2j+2}\right)
+\varepsilon
_{b,2j+1}b_{2j+1}+U_{bb}|b_{2j+1}|^{2}b_{2j+1}+ \nonumber \\
&& + U_{ab}\left( |a_{2j}|^{2}+|a_{2j+2}|^{2}\right) b_{2j+1},
\label{eq1b}
\end{eqnarray}
where $\Delta \omega =(\omega _{1}-\omega_{2})$, the parameters
$U_{aa},U_{bb}$ define interaction between atoms induced by on-site
atomic collisions and interaction between atoms of neighbor sites is
defined by $ U_{ab}$.

The first term in these equations takes into account the nearest-
neighbor hopping induced by the stimulated Raman scattering. We
assume that inhomogeneous broadening is absent, i.e., $\varepsilon
_{a,2j}=\varepsilon _{a}$ and $\varepsilon
_{b,2j+1}=\varepsilon_{b}$. If we introduce the control
electromagnetic field amplitudes $\mathcal{E}_{1.2}$ then the
nearest-neighbor hopping term read as $J_{0}=\mu
_{12}\mathcal{E}_{1}\mathcal{E}_{2}^{\ast }$, where $\mu_{12} $ is
the matrix element of the Raman transition. If one introduce
$J_{0}=|J_{0}|\exp (i\vartheta )$, and assume that control
electromagnetic fields have a constant phase, then $ \vartheta$ can
be included into complex value of the probability amplitudes
$b_{2j+1}$. Thus we can substitute $ b_{2j+1}\rightarrow b_{2j+1}
\exp(i\vartheta) = \tilde b_{2j+1}$, after that suppose the
parameter $J_{0}$ as real value. If the interaction between atoms of
neighbor sites is neglected then the system of resulting equations
takes the following form
\begin{equation}
\begin{array}{rcl}
i\partial \tilde{a}_{2j}/\partial \tau & = & -\left(
\tilde{b}_{2j-1}+\tilde{
b}_{2j+1}\right) +\beta _{a}|\tilde{a}_{2j}|^{2}\tilde{a}_{2j}, \\
i\partial \tilde{b}_{2j+1}/\partial \tau & = & -\left(
\tilde{a}_{2j}+\tilde{ a}_{2j+2}\right) +\delta
\tilde{b}_{2j+1}+\beta _{b}|\tilde{b}_{2j+1}|^{2} \tilde{b}_{2j+1},
\end{array}
\label{eq5}
\end{equation}
where $\Delta \varepsilon =(\varepsilon _{b}-\varepsilon _{a})-\hbar
\Delta \omega $, $\beta _{a}=U_{aa}/|J_{0}|$, $\beta
_{b}=U_{bb}/|J_{0}|,$, $ \delta =\Delta \varepsilon /|J_{0}|$. We
use the normalized time variable $\tau =t|J_{0}|/\hbar $. For the
sake of simplicity we will assume that the exact resonance condition
is hold, i.e., $ \delta=0$.

\section{Stationary solution }

It should remark that the atomic transport between neighbor sites in
superlattice is absent if the phases of amplitudes $a_{2j}$  and
$a_{2j+2}$ as well as  $b_{2j-1}$  and $b_{2j+1}$ will be opposite
one.

Fig. 1 represents schematically the probability amplitudes
configurations in superlattice. Thin line arrows correspond to even
sites the twin-line arrows correspond to odd sites. Sing of the
probability amplitude is indicated by orientation of the arrow,
i.e., plus (minus) corresponds to directed up (down) arrow. The
configurations shown in Fig.1 (a) and (b) are characterized by same
energy, hence we can except the existence of the solution of the
equations (\ref{eq5}) which describe the domain wall separated these
two configurations. However, there we will not consider this case.

The insertion of the ansatz  $\tilde{a}_{2j}=(-1)^{j}a(\tau )$ and
$\tilde{b}_{2j+1}=(-1)^{j}b(\tau )$ into equations (\ref{eq5})
results in following system of equations for the probability
amplitudes
\begin{equation}
i\frac{\partial a}{\partial \tau }=\beta _{a}|a|^{2}a,\quad i\frac{\partial b%
}{\partial \tau }=\beta _{b}|b|^{2}b  \label{eq7}
\end{equation}
Equations (\ref{eq7}) show that populations of the sites of each
sublattice are independent. It is convenient rewrite the equations
(\ref{eq7}) in term of real variables
\[
a(\tau )=u(\tau )\exp \{\varphi _{a}(\tau )\},\quad b(\tau )=w(\tau
)\exp \{\varphi _{b}(\tau )\}
\]
that leads to the system of simple real equations
\[
\frac{\partial u}{\partial \tau }=0,~\frac{\partial w}{\partial \tau
}=0, ~\frac{\partial \varphi _{a}}{\partial \tau }=-\beta
_{a}u^{2},~\frac{\partial \varphi _{b}}{\partial \tau }=-\beta
_{b}w^{2}.
\]
Solutions of these equations read as
\begin{equation}
u(\tau )=u_{0},~w(\tau )=w_{0},~\varphi _{a}(\tau )=\varphi
_{a0}-\beta _{a}u_{0}^{2},~\varphi _{b}(\tau )=\varphi _{b0}-\beta
_{b}w_{0}^{2} \label{eq9}
\end{equation}
Choosing of the initial phases we can state the configuration of the
initial probability amplitude distribution as it shown in Fig.1
(a)($\varphi _{a0}= \varphi _{b0}=0$), or in Fig.1(b) ($\varphi
_{a0}=0,\varphi _{b0}=\pi $).

\section{Stability analysis for stationary distribution }

Stability of the solution found above will be analyzed in the
framework of the linear stability theory. Let us consider the small
perturbations of the stationary distribution
\begin{equation}
\tilde{a}_{2j}=(-1)^{j}a(\tau )+\delta
a_{2j},~~\tilde{b}_{2j+1}=(-1)^{j}b( \tau )+\delta b_{2j+1},
\label{eq10b}
\end{equation}
with
\[
a(\tau )=u_{0}\exp \{-\beta _{a}u_{0}^{2}\tau \},~~b(\tau
)=w_{0}\exp \{-\beta _{b}w_{0}^{2}\tau \}
\]
The initial phases are chosen in the following form: $\varphi
_{a0}=0,\varphi _{b0}=0$.

The linear equations associated with (\ref{eq5}) read as
\begin{equation}
i\frac{\partial }{\partial \tau }\delta a_{2j}=-\left( \delta
b_{2j-1}+\delta b_{2j+1}\right) +2\beta _{a}u_{0}^{2}\delta
a_{2j}+\beta _{a}a^{2}\delta a_{2j}^{\ast },  \label{eq12a}
\end{equation}
\begin{equation}
i\frac{\partial }{\partial \tau }\delta b_{2j+1}=-\left( \delta
a_{2j}+\delta a_{2j+2}\right) +2\beta _{b}w_{0}^{2}\delta
b_{2j+1}+\beta _{b}b^{2}\delta b_{2j+1}^{\ast },  \label{eq12b}
\end{equation}
If one substitute $\delta a_{2j}(\tau )=p_{2j}(\tau )\exp \{-\beta
_{a}u_{0}^{2}\tau \}$, $\delta b_{2j+1}(\tau )=q_{2j+1}(\tau )\exp
\{-\beta _{b}w_{0}^{2}\tau \}$, than (\ref{eq12a}) and (\ref{eq12b})
can be rewritten as
\[
i\frac{\partial }{\partial \tau }p_{2j}=-\left(
q_{2j-1}+q_{2j+1}\right) \exp \{i(\varphi _{b}-\varphi _{a})\}+\beta
_{a}u_{0}^{2}\left( p_{2j}+p_{2j}^{\ast }\right) ,
\]
\[
i\frac{\partial }{\partial \tau }q_{2j+1}=-\left(
p_{2j}+p_{2j+2}\right) \exp \{i(\varphi _{a}-\varphi _{b})\}+\beta
_{b}w_{0}^{2}\left( q_{2j+1}+q_{2j+1}^{\ast }\right).
\]
Assume that the constant probability amplitudes (or the site
population of superlattice) are related by the following expression
\begin{equation}
\beta _{a}u_{0}^{2}=\beta _{b}w_{0}^{2}=\lambda _{1}.  \label{eq14}
\end{equation}
In this case the phase difference $\varphi _{b}-\varphi _{a}$ will
be constant. We can put it to zero. Thus, the system of linear
equations for small perturbations takes the form
\begin{equation}
\begin{array}{lcl}
i\partial p_{2j}/\partial \tau  & = & -\left(
q_{2j-1}+q_{2j+1}\right)
+\lambda _{1}\left( p_{2j}+p_{2j}^{\ast }\right) , \\
i\partial q_{2j+1}/\partial \tau  & = & -\left(
p_{2j}+p_{2j+2}\right) +\lambda _{1}\left( q_{2j+1}+q_{2j+1}^{\ast
}\right) .
\end{array}
\label{eq15m}
\end{equation}
Substitution of the following expressions
\begin{equation}
\begin{array}{lcl}
p_{2j} & = & A\exp (2ijkl)+B\exp (-2ijkl) \\
q_{2j+1} & = & C\exp \{i(2j+1)kl\}+D\exp \{-i(2j+1)kl\}
\end{array}
\label{eq16}
\end{equation}
into the differential-difference equations (\ref{eq15m}) leads to
the system of linear differential equations
\[
\begin{array}{cccccc}
i\partial A/\partial \tau & = & -2\cos klC+\lambda _{1}\left(
A+B^{\ast }\right) , & ~~i\partial A^{\ast }/\partial \tau & = &
2\cos klC^{\ast
}-\lambda _{1}\left( A^{\ast }+B\right) , \\
i\partial B/\partial \tau & = & -2\cos klD+\lambda _{1}\left(
A^{\ast }+B\right) , & ~~i\partial B^{\ast }/\partial \tau & = &
2\cos klD^{\ast
}-\lambda _{1}\left( A+B^{\ast }\right) , \\
i\partial C/\partial \tau & = & -2\cos klA+\lambda _{1}\left(
D^{\ast }+C\right) , & ~~i\partial C^{\ast }/\partial \tau & = &
2\cos klA^{\ast
}-\lambda _{1}\left( C^{\ast }+D\right) , \\
i\partial D/\partial \tau & = & -2\cos klB+\lambda _{1}\left(
C^{\ast }+D\right) , & ~~ i\partial D^{\ast }/\partial \tau & = &
2\cos klB^{\ast }-\lambda _{1}\left( D^{\ast }+C\right) .
\end{array}
\]
It is convenient introduce new variable $\xi =\lambda _{1}\tau $ and
constant parameter  $\mu =2\cos kl/\lambda _{1}=2\cos kl/\beta
_{a}u_{0}^{2}$. From the foregoing equations one can obtain the
system of equations of second order
\begin{equation}
\begin{array}{rclrcl}
\partial ^{2}A/\partial \xi ^{2} & = & -\mu ^{2}A+2\mu C, & \partial
^{2}C/\partial \xi ^{2} & = & -\mu ^{2}C+2\mu A, \\
\partial ^{2}B/\partial \xi ^{2} & = & -\mu ^{2}B+2\mu D, & \partial
^{2}D/\partial \xi ^{2} & = & -\mu ^{2}D+2\mu B.
\end{array}
\label{eq17}
\end{equation}
Now, the characteristic equation for this system of equations
(\ref{eq17}) can be determined easily
\begin{equation}
\Upsilon(\sigma)= \mathrm{Det}\left(
\begin{array}{cccc}
\sigma ^{2}-\mu ^{2} & 2\mu & 0 & 0 \\
2\mu & \sigma ^{2}-\mu ^{2} & 0 & 0 \\
0 & 0 & \sigma ^{2}-\mu ^{2} & 2\mu \\
0 & 0 & 2\mu & \sigma ^{2}-\mu ^{2}
\end{array}
\right) =0  \label{eq19}
\end{equation}
Stability of the solutions of the equations (\ref{eq9}) is
determined by the roots of this equation, which can be written as
\[
\sigma _{\pm }^{2}=\mu ^{2}\pm 2|\mu |=\left( |\mu |\pm 1\right)
^{2}-1.
\]
Instability of the configuration of site population under
consideration means that imaginary part of the any root is not zero.
But if $\sigma _{\pm }^{2}$ is positive one, then Im$\sigma =0$. One
should note that $\sigma _{+}^{2}\geq 0$ for any $ |\mu |$,  whereas
$\sigma _{-}^{2}\geq 0$ only at $ |\mu |\geq 2$. Hence, one can
conclude that the configuration of site population is stable under
following condition
\begin{equation}
|\cos kl|\geq \beta _{a}u_{0}^{2}  \label{eq21}
\end{equation}
In terms of physical meaning variable this inequality is read as
\begin{equation}
U_{bb}w_{0}^{2}=U_{aa}u_{0}^{2}\leq |J_{0}||\cos kl|.  \label{eq22}
\end{equation}
Else, the modulation instability takes place if
\begin{equation}
|\cos kl| <\beta _{a}u_{0}^{2}  \label{eq23}
\end{equation}
The amplitude of small perturbations varies as $\exp (i\sigma _{\pm
}\xi )= \exp (i\sigma _{\pm }\lambda _{1}\tau )$. As it was
indicated above exponential growing of the amplitude is related with
parameter $\sigma _{-}\lambda _{1}$. The imaginary part of $\sigma
_{-}\lambda _{1}$ is the instability increment $G(k)$, i.e., :
\begin{equation}
G^{2}(k)=4|\cos kl|\left( \beta _{a}u_{0}^{2}-|\cos kl|\right)
\label{eq24}
\end{equation}
If we consider the first Brillouin zone $-\pi /2\leq kl\leq \pi /2$,
then the stability region lies into interval $-\arccos (\beta
_{a}u_{0}^{2})<kl<\arccos (\beta _{a}u_{0}^{2})$. The instability
regions are determined by the inequalities $-\pi /2<kl<-\arccos
(\beta _{a}u_{0}^{2})$, $\arccos (\beta _{a}u_{0}^{2})<kl<\pi /2$.
One can found that the instability increment is zero at boundary
points of these regions. Maximum of the increment placed at points
$k_{m}l=\pm \arccos (\beta _{a}u_{0}^{2}/2)$ and maximum magnitude
of increment is equal to $G_{m}=G(k_{m})=\beta _{a}u_{0}^{2}$. And
at $\beta _{a}u_{0}^{2}>2$ the increment, ones taken at point
$k_{m}l=0$, has maximum value $G_{m}=2\sqrt{\beta _{a}u_{0}^{2}-1}$.

Fig.2 show the dependence of the instability increment on wave
number of the weak perturbation and on nonlinearity parameter
$\lambda _{1}=\beta _{a}u_{0}^{2}=\beta _{b}w_{0}^{2}$. This
parameter is defined by the population of the superlattice sites. As
one can see increasing of the $\lambda _{1}$ results in decreasing
of the stability region. At $\lambda _{1} \geq 1$ the instability
region occupies  first Brillouin zone totally. All solutions
(\ref{eq9}) are instable.

\section{Numerical analysis }

The aim of numerical simulation is study of the evolution of
stationary configuration of the site population found above in
response to a weak harmonic perturbation. Throughout this simulation
relation $\beta _{a}u_{0}^{2}=\beta _{b}w_{0}^{2}$ is assumed. The
equations (\ref{eq5}) with $\delta = 0$ were solved at the following
initial ($\tau=0$) conditions
\[
\begin{array}{ccl}
\tilde{a}_{2j}(0) & = & (-1)^{j}u_{0}\exp \{-i \lambda_{1} \tau \} +
\delta a \cos(kl(2j)) \\
\tilde{b}_{2j+1}(0) & = & (-1)^{j}w_{0}\exp \{-i \lambda_{1} \tau \}
+ \delta a \cos(kl(2j+1))
\end{array}
\]
There we consider the periodic boundary condition: $b_{n+1} =
b_{1}$, $a_{n+2} = a_{2}$, where $n$ is total number of sites in
superlattice. The superlattice length was chosen to be multiple of
the half-period of perturbation.

As example we represent results of the numerical simulation of the
modulational instability in superlattice containing $400$ sites,
where the initial values for probability amplitudes are $a(0) = 2$,
$b(0) = 3$, perturbation amplitude is $\delta a = 0.01$ and
perturbation wave number is $k = 0.039/l$ ($l$ is distance between
neighbor sites). If nonlinearity parameter $\lambda_{1} = \beta
_{a}u_{0}^{2}$ is over one the modulational instability manifests
itself causing an exponential growth of small perturbations of the
harmonic wave (Fig.3) and (Fig.4). By using the initial value
probability amplitude of population for site of $a$ -type, (i.e.,
$a(0)$) and the same value at time $\tau $, i.e. $a(\tau)$, one can
calculate the instability increment according to formula $G_{num} =
\ln[(a(\tau) -a(0))/\delta a]/\tau $ (the same we can done for site
of $b$ -type). On the other hand, value of the increment $G$ is
determined by formula (\ref{eq24}). For  $\beta _{a}u_{0}^{2} = 2.5$
(that corresponds to $\beta _{a} = 0.625$ and $\beta_{a} = 0.277$)
we obtained $G_{num} = 2$ and $G = 2.4$. If the value $\beta
_{a}u_{0}^{2}$ is descried up to one under condition that all
parameters of system are fixed, instability persists. However, the
instability increment reduces progressively downstream. For $\beta
_{a}u_{0}^{2} < 1$ the regions of stability and instability
appeared. Thus we can conclude that instability state of the site
population is typical for high populations of the sites in
superlattice and for the case, where an on-site interaction between
bosons dominates over photo-induced transport. Otherwise one can
observe stable state picture: oscillation of an excess population
near stable value, as shown in Fig.5.

Now we consider dynamics of the perturbations with different spatial
frequencies. The nonlinearity parameter $\beta_{a}u_{0}^{2}$ , which
put to be less than one, is fixed, but wave number of the harmonic
perturbation $k$ (see expression for initial conditions) will be
varied. We can expect transition from stability to instability at
$kl = \arccos(\beta_{a}u_{0}^{2})$ because of the stability region
lies into interval $-\arccos (\beta _{a}u_{0}^{2})<kl<\arccos
(\beta_{a}u_{0}^{2})$. Put $\beta _{a}u_{0}^{2}= 0.99$ ($\beta _{a}
= 0.2475$, $\beta _{b}  = 0.11$). Fig.6 and Fig.7 represent the time
dependences of the site populations (i.e., square of modulus of
probability amplitudes) respectively for $k_{1}l = 0.10244$
~$(\cos(k_{1}l) = 0.9948 )$ and $k_{1}l = 0.18124$ ~$(\cos(k_{1}l) =
0.9836$).

\section{Conclusion}

In this work we have studied Bose-Einstein condensates in the one
dimensional optical superlattice with two kinds of microscopic
potentials (sites of the lattice). It was assumed that the deep of
these potentials is enough the system to be described by the
Bose-Habbard model. Steady state distribution of the probability
amplitudes of site population was found on base of the earlier
derived system of equations (\ref{eq5})~\cite{R13}, determining the
dynamic of probability amplitude of sites filling (or the
probability to find the boson in this site) in superlattice. Feature
of this stationary state is the phase alternating, i.e., phases of
probability amplitude of one-type sites change on $\pi$ while going
from one site to another. The stability of such state of BEC for
small perturbations depends on the nonlinearity parameter $\beta
_{a}u_{0}^{2}$ and also on the frequency of modulation by itself.
When the value of $\beta _{a}u_{0}^{2}$ exceeds one, the found
distribution is unstable for all wave numbers from Brillouin zone.
However, when $\beta _{a}u_{0}^{2}$  is less than one the
instability region in the Brillouin zone are determined by
inequalities  $-\pi /2<kl<-\arccos (\beta _{a}u_{0}^{2})$ and
$\arccos (\beta _{a}u_{0}^{2})<kl<\pi /2$. So, the bosons
distribution is modulationaly instable for short wavelength
perturbation, whereas the long wavelength perturbations dump out. If
the nonlinearity parameter (or the average number of bosons in one
site of superlattice) decreases, the instability region reduces. The
values of the instability increment found by analytically were
compared with the results of numerical simulation, and good
agreement between them was found in the field, where the linear
analysis of stability is valid.

The nonlinear regime of the modulational instability was studied by
using the numerical simulation. It was shown that the number of
maxima in population distribution on sites appears. It should be
remarked that the preliminary calculations demonstrate the
development of chaotic behavior of the considered system.

It is necessary to notice that in ordinary optical lattice the
equations of motion for the probability amplitudes of the sites
population in continual limit may be transformed into the nonlinear
Schr\"{o}dinger equation having soliton solutions. It describes
approximately the spatial solitons in ordinary optical lattices. In
case under considering here the equations of motions (\ref{eq5}) in
continual limit result in more complex equations, which are not like
to be completely integrable. So it is not necessary to expect
formation of solitons chain as result of modulational instability.
Inelastic interaction between solitary spatial waves appearing
there, likely will leads to chaotic bosons distribution per sites.

Finally, we assume it not required an especially effort for
generalisation of the present results on the 2D cases, e.g., for the
simple square or cubic superlattice.

\section*{Acknowledgment}

We are grateful to S.O. Elyutin for valuable discussions. The work
was supported by the Russian Basic Research Foundation (Grant No
06-02-16406).

\newpage

\section*{FIGURE CAPTIONS}

Fig.~1. Two allowed configurations of the probability amplitudes
distribution in superlattice (a) $\varphi _{a0}=\varphi _{b0}=0$,
(b) $\varphi _{a0}=0, \varphi _{b0}=\pi. $.

Fig.~2. Instability increment $G(k)$ versus $k$ and $\lambda _{1}$.

Fig.~3. Strong modulational instability. Evolution of the site
populations in superlattice for nonlinearity parameter $\beta
_{a}u_{0}^{2}= 2.5$. .

Fig.~4. Evolution of the sites populations for nonlinearity
parameter $\beta {a}u_{0}^{2}= 2.5$. Solid line corresponds to site
with maximum of value of initial perturbation, dash line corresponds
to site with maximum of negative value of initial perturbation, dot
lines correspond to sites with intermediate value of the initial
perturbations.

Fig.~5. Same as in Fig.4 but for nonlinearity parameter $\beta
{a}u_{0}^{2}= 0.8$. and the dot lines correspond to sites with
minimum value of initial perturbation.

Fig.~6. Dynamic of weak harmonic perturbation with $k_{1}l=0.10244$,
of the stability region..

Fig.~7. Same as in previous figure, but for harmonic perturbation
with $k_{1}l = 0.18124$. One can see, that the distribution is
instable now ($G_{num} = 0.08$, $G = 0.14$).

\end{document}